%% file: co2ode_main.tex
\documentclass{article}

\PassOptionsToPackage{numbers,sort&compress}{natbib}
\usepackage[dblblindworkshop, final]{neurips_2026}
\workshoptitle{Sim2Science: ML with Imperfect Scientific Models}

\usepackage[utf8]{inputenc} 
\usepackage[T1]{fontenc}    
\usepackage{amsmath}
\usepackage{hyperref}       
\hypersetup{hidelinks}      
\usepackage{url}            
\usepackage{booktabs}       
\usepackage{float}          
\usepackage{amsfonts}       
\usepackage{nicefrac}       
\usepackage{microtype}      
\usepackage{xcolor}         
\usepackage{graphicx}
\graphicspath{{figures/}}

\title{Co\texorpdfstring{\textsuperscript{2}}{2}ODE: conditionally coarse-grained model discovery for rewiring in biochemical networks}

\author{%
  Theodore de Pomereu\textsuperscript{1,2} \qquad Fabian Fröhlich\textsuperscript{1} \\
  \textsuperscript{1}Dynamics of Living Systems Laboratory, The Francis Crick Institute, London, UK \\
  \textsuperscript{2}Department of Mathematics, Imperial College London, London, UK \\
  \texttt{\{theodore.pomereu, fabian.frohlich\}@crick.ac.uk}
}

\begin{document}

\maketitle

\begin{abstract}
Biochemical networks propagate the signals that regulate cellular behaviour, and which routes carry them depends on cellular context: a shift in context rewires the network's effective dynamics. Mechanistic simulators are misspecified here because the effective dynamics emerge from an interaction structure that is only partially known. Model discovery instead learns the dynamics from data, but sparse, correlated measurements poorly identify a full network model. What matters is narrower: the portion the context rewires. We introduce conditionally coarse-grained ODEs (Co\textsuperscript{2}ODE), which model that portion as a sparse neural or symbolic ODE: a conditional mask selects the markers carrying the rewiring, and a variational autoencoder compresses the remaining markers into a learnt closure term. On single-cell phosphoproteomic data spanning 30 cancer-relevant overexpressions, Co\textsuperscript{2}ODE compresses a 30-marker panel into a 10-dimensional state while matching the predictive accuracy of a neural ODE fit to the full panel. The closure absorbs what the mask drops, hence the accuracy is insensitive to which markers are kept. Adding a prior score to the loss shifts the selected markers at no cost in fit, indicating the data are not interventional enough to identify the underlying biology.
\end{abstract}

\section{Introduction}

Biochemical networks underlie most cellular regulation, and their dysregulation underpins many diseases, cancer in particular~\cite{alberts2015molecular,hanahan2011hallmarks}. Dysregulation arises when a network's topology or its surrounding context is perturbed: protein abundances and feedback strengths shift, the effective interaction strengths change, and with them how signal propagates~\cite{lun2019kinome}. The path to a target node is then rerouted or interfered with by alternative pathways, what we call rewiring. It limits targeted therapy for cancer: inhibiting a protein is a known change, but the rewiring that follows is not, and signal can reach the target by a route the inhibitor does not block~\cite{prahallad2012unresponsiveness,chandarlapaty2012negative}, differently in each patient.

Modelling rewiring means recovering, for each context, which interactions carry the signal, and what shape their dynamics take. Hand-built mechanistic simulators encode prior knowledge and are fit to data~\cite{frohlich2018efficient}, but incomplete knowledge and poor transferability between contexts limit their predictiveness, and their scale leaves them weakly identifiable from partial, noisy measurements~\cite{raue2009structural}. Data-driven model discovery avoids relying on prior knowledge entirely~\cite{schmidt2009distilling,chen2018neural}, but its difficulty grows with the dimensionality of intracellular networks. Existing methods for discovering rewiring learn interaction structure from perturbation response~\cite{yuan2021cellbox,kholodenko2002untangling}, but they assume the intervention acts on a node the model contains and the system sits at steady state.

In this study, we learn the coarse-graining and the dynamics simultaneously, removing the need for a trusted simulator or a full-dimensional learnt model. This is tractable because we do not need a general coarse-graining, only the markers carrying signal to the target in a given context. \textit{Conditionally coarse-grained ODE} (Co\textsuperscript{2}ODE) is an end-to-end differentiable architecture that learns such a reduced model directly from data.

\textbf{Our contributions} are threefold. First, we learn a context-conditional coarse-graining jointly with the dynamics, such that a reduced model can be found where no parent model exists to reduce. Second, we build two prior-knowledge scores that evaluate the recovered structure against a curated network. These scores can also be added to the model as inductive biases. Our third contribution is to apply the model to real signalling data where we find that the node score relocates that structure at no cost in fit, showing these data do not identify the true rewiring, and agreement with known biology cannot validate what such a model recovers.

\section{Related Work}
\label{sec:related}

\textbf{Coarse-graining biochemical networks.} There are many different approaches to model reduction in the literature: time-scale separation~\cite{segel1989quasi,goeke2017classical}, lumping~\cite{okino1998simplification}, pruning~\cite{snowden2017methods}, profile-likelihood reduction~\cite{maiwald2016driving}, manifold boundary approximation~\cite{transtrum2014model}, and projection methods~\cite{moore1981principal,liebermeister2005biochemical,budisic2012applied}. Most often they reduce an already existing parent model, and preserve their behaviour under conditions prescribed in advance. Here there is no such model, and the behaviour of interest is what changes with context.

\textbf{Model discovery.} This approach infers dynamics from data with little or no prior structure, from symbolic regression to neural ODEs~\cite{schmidt2009distilling,chen2018neural,rackauckas2020universal,raissi2019physics,yazdani2020systems}, with sparsity regularisation giving interpretable structure in the learnt network~\cite{aliee2022sparsity,aliee2021beyond}. However the symbolic search space grows combinatorially with input dimension and neural formulations remain weakly identified in the same regime as the simulators they replace.

\textbf{Closure.} Removing variables from a dynamical system destroys Markovian closure: the retained variables no longer evolve autonomously, which can dampen predictiveness. The Mori--Zwanzig formalism restores this information by constructing a memory integral over the history of the retained variables together with a fluctuating remainder~\cite{mori1965transport,zwanzig1961memory,chorin2002optimal}, and recent work learns it from data instead of deriving it~\cite{ma2019model,gupta2021neural}. In our case, the discarded markers enter the model through the present-time latent evolved alongside the reduced state; the closure is therefore Markovian.

\textbf{Modelling rewiring from perturbation data.} Modular-response analysis recovers
connection strengths from steady-state responses to module-specific
perturbations~\cite{kholodenko2002untangling}, applied per cell line to quantify EGFR
signalling and predict resistance-breaking combinations~\cite{klinger2013network}. cSTAR
pairs that inference with a supervised axis between known cell states, reducing to a core
circuit that differs between cell lines and between receptor variants of one
line~\cite{rukhlenko2022control}. These approaches need a perturbation specific to each
module, and the modules are prescribed in advance and shared across contexts. CellBox
instead learns a full interaction matrix at panel
dimension~\cite{yuan2021cellbox,ji2021inference}. Our contexts act from outside the measured
state and are seen only through a reporter, the data are out-of-equilibrium time courses, and the
reduction is learnt per context rather than fixed.

\section{Methods}
\label{sec:method}

\begin{figure}[t]
  \centering
  \includegraphics[width=0.995\linewidth]{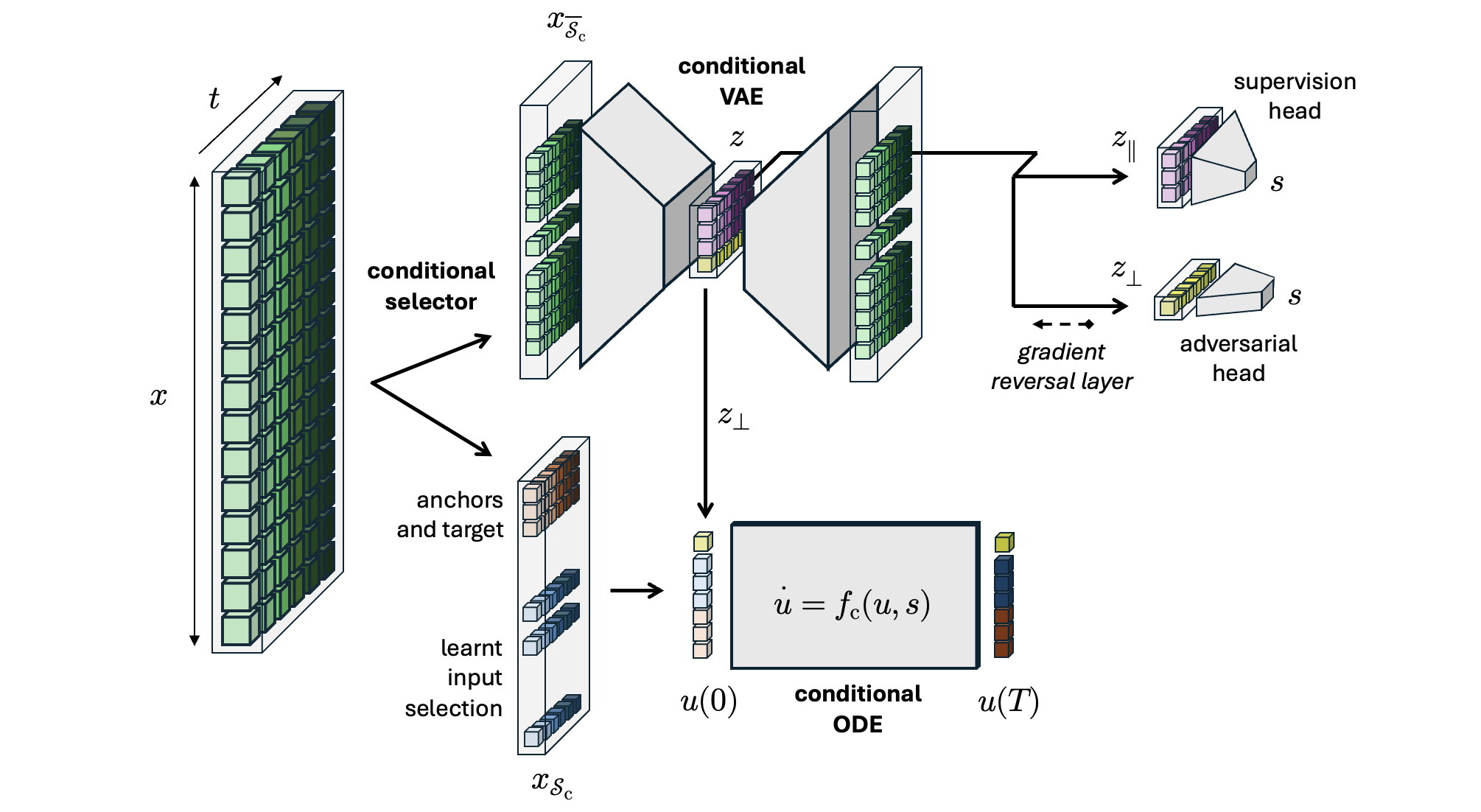}
  \caption{Co\textsuperscript{2}ODE. A context-conditional selector keeps the markers that carry the rewiring. The discarded markers are compressed into a latent closure, which joins the selected markers to form the coarse state. That state evolves under a context-conditioned sparse neural or symbolic ODE.}
  \label{fig:arch}
\end{figure}

Co\textsuperscript{2}ODE takes trajectories from a biochemical panel measured across
contexts and returns, per context, a low-dimensional ODE around a designated target node. The architecture has three components, trained jointly (Figure~\ref{fig:arch}). A
context-conditional \emph{selector} chooses which measured markers enter the model.
The markers left out are handled by a \emph{latent closure}. The retained markers
and the latent are then evolved by a \emph{sparse dynamics} module, conditioned on
context and realised either as a neural or a symbolic ODE.

\subsection{Problem setup}
\label{sec:setup}

For each context $c$ we observe a panel of $M$ markers,
$x^{c,s}(t)\in\mathbb{R}^{M}$, over $t\in[0,T]$. The scalar $s\in[0,1]$ is the
strength at which the context is induced. Each context is measured at a finite number of strengths,
and we assume $x^{c,s}$ varies smoothly in $s$, so intermediate strengths
interpolate between the levels observed. One designated coordinate
$y^{c,s}(t)=x^{c,s}_{r}(t)$ is the target. We want to learn a small closure around
$y^{c,s}$: a reduced state $u^{c,s}\in\mathbb{R}^{d}$, $d\ll M$, comprising the
target and a few other coordinates, evolving as
$\dot u^{c,s}=f_c(u^{c,s},s)$. Contexts differ in which coordinates are selected
and in the field $f_c$. That difference is the rewiring.

\subsection{Context-conditional state selection}
\label{sec:selector}

We select the markers entering the reduced state in two parts. We always keep
the target $r$ and a small anchor set $\mathcal{A}\subset[M]$, the canonical
routes to the target. The rest are chosen by a straight-through top-$k$ ($k=3$)
over per-context logits $\ell_{c,i}$, $i\in[M]\setminus(\{r\}\cup\mathcal{A})$.
Forward, the hard mask $m_c\in\{0,1\}^{M}$ keeps the $k$ largest; backward,
gradients flow through $\mathrm{softmax}(\ell_c/\tau)$, where $\tau$ is the
annealing temperature~\cite{bengio2013estimating}. Masks are indexed by context
alone so a single selection applies across the context strength gradient. Thus $\mathcal{S}_c=\{r\}\cup\mathcal{A}\cup\{i:m_{c,i}=1\}$ has
$1+|\mathcal{A}|+k$ elements, only $k$ of them learned. Throughout,
$x_{\mathcal{I}}$ denotes the subvector on an index set $\mathcal{I}$, entries
in index order, such that the selected markers are $x^{c,s}_{\mathcal{S}_c}$ and
the discarded markers are $x^{c,s}_{\bar{\mathcal{S}}_c}$, with
$\bar{\mathcal{S}}_c=[M]\setminus\mathcal{S}_c$.

\subsection{Closure by a latent}
\label{sec:closure}

An ODE over the selected markers alone is not necessarily closed: The discarded markers could still
act on them, and the ODE has no term for that influence. We therefore encode
$x^{c,s}_{\bar{\mathcal{S}}_c}$ into a latent $z^{c,s}$ with a conditional
variational autoencoder (CVAE)~\cite{kingma2014auto,sohn2015learning}, and split it into
$z^{c,s}_{\parallel}$ and $z^{c,s}_{\perp}$ to discourage it from becoming a second
route for the context signal. We achieve this by regularising the CVAE with two heads. One of them predicts $s$ from $z^{c,s}_{\parallel}$, while a gradient-reversal
layer~\cite{ganin2015unsupervised} trains the encoder so that $s$ is not predictable
from $z^{c,s}_{\perp}$. Only $z^{c,s}_{\perp}$ joins the selected markers, giving the
coarse state $u^{c,s}(t)=[\,x^{c,s}_{\mathcal{S}_c}(t),\,z^{c,s}_{\perp}(t)\,]$: a
learned Markovian closure~\cite{gupta2021neural} whose latent supplies background
dynamics.

\subsection{Context-conditioned dynamics}
\label{sec:dynamics}

The coarse state evolves under $\dot u^{c,s}=f_c(u^{c,s},s)$ and is integrated with a fixed-step Runge--Kutta scheme to predict the target trajectory. The vector field $f_c$ is learnt in one of two ways: (i) as a \emph{sparse neural ODE} carrying a group-sparsity (L21) penalty on its Jacobian (Appendix~\ref{si:limits}), or (ii) a \emph{conditional SINDy} module~\cite{brunton2016discovering} that regresses over a non-linear function library sparsely with per-context coefficients. In both cases, the strength modulates the coefficients continuously. The rewiring structure is read off the Jacobian elements $\partial f_{c,i}/\partial u^{c,s}_j$. We report the neural implementation in the remainder of the paper unless stated otherwise; the symbolic version is discussed in Appendix~\ref{si:symbolic}.

\subsection{Training}
\label{sec:training}

All components are differentiable and are trained jointly against a weighted sum of the mean squared error of the target trajectory integrated from $t=0$, the autoencoder's negative ELBO (Gaussian reconstruction and KL to a standard normal prior), the mean-squared supervised and adversarial strength regressions, and the dynamics sparsity penalty. The neural variant trains in one stage with Adam~\cite{kingma2015adam}; the SINDy variant adds a second stage that freezes the selector and autoencoder and sparsifies the dynamics. Schedules are annealed over training; hyperparameters are in Appendix~\ref{sec:appendix}.

\section{Scoring the model against prior knowledge}
\label{sec:prior}

Our prior is a curated directed interaction network over the markers and overexpressed genes, built with
OmniPath~\cite{turei2016omnipath}. The \emph{node score} charges each selected
marker for the detour it forces: for a marker $i$, we compare the shortest path
from the perturbed node $p$ to the target $r$ that passes through $i$ against the
shortest path overall. The \emph{edge score} penalises each dependency the model learns---a nonzero
Jacobian entry, or an active symbolic term---by the shortest-path distance between
the two markers on the prior network. Direct edges are free; longer paths cost more,
and pairs with no path at all cost the most. Both the prior and the distance are
given in Appendix~\ref{si:prior}.

\section{Experiments}
\label{sec:experiments}

\subsection{Data, metrics, and baselines}

We use the single-cell mass-cytometry dataset from Lun \textit{et al.} (2019)~\cite{lun2019kinome}, in which single-gene overexpression induces a set of contexts in a common cell background. A 30-marker phospho panel is read out in 6 snapshots over one hour, following stimulation upstream of the network by the growth factor EGF.
Two conditions are dropped for mismatched time grids, leaving 38: the 30 overexpressions we report on, and 8 controls that carry no overexpression. Trajectories are Gaussian-process smoothed~\cite{rasmussen2006gaussian}. The context strength is a GFP-derived proxy for induction, the target is p-S6 (a growth-signalling readout), and the anchors are p-MEK and p-AKT (upstream pro-proliferation nodes); the $80/10/10$ split gives $2919/365/365$ (Appendix~\ref{si:data}).

We report range-normalised mean absolute error (rnMAE) on the target trajectory from $t=0$; the cross-seed overlap of the selected marker set; and the two prior scores of Section~\ref{sec:prior} (Appendices~\ref{si:metric},~\ref{si:stats}).

Two ablation studies each change one part of the model. The first varies how much of the panel enters the state: anchors and target alone (3), plus the markers the selector adds (6), plus the four-dimensional closure (10, Co\textsuperscript{2}ODE), and the panel unreduced (30). The 3- and 6-dimensional arms are closure-free, not latent-free: they train the same autoencoder, but its latent never enters the ODE. The second varies how the biases enter: not at all, node score, edge score, or both. Every arm is swept over the same grid and selected on validation (Appendices~\ref{si:selection},~\ref{si:compute}).

\begin{figure}[t]
  \centering
  \setlength{\abovecaptionskip}{4pt}
  \includegraphics[width=\linewidth]{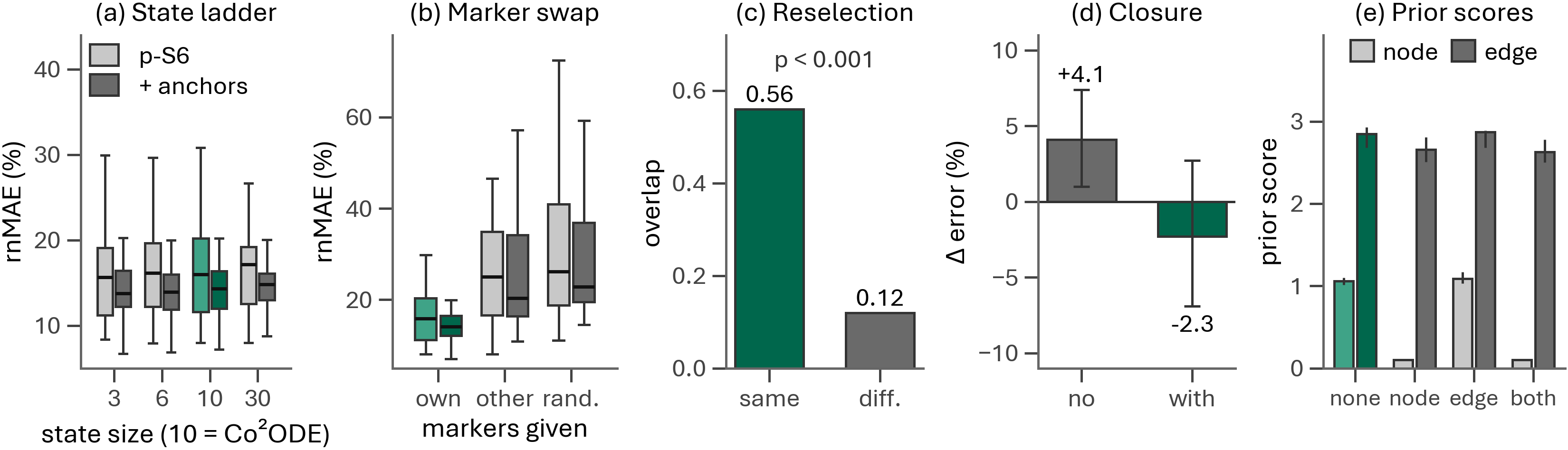}
     \caption{rnMAE (Appendix~\ref{si:metric}); lower is better. Boxes in (a, b) show the 38
    per-context values, each a median over seeds (10 per arm; 5 for the 30-marker arm); whiskers
    $1.5\times$IQR. (a) Reducing the state from 30 to 10, 6 or 3 variables costs at most half a
    point. (b) Swapping in another context's or random markers costs about ten points. (c) Seed-to-seed
    Jaccard overlap of selections, same vs.\ mismatched context; permutation test. (d) Extra error of
    runs departing from the modal selection, with and without closure; bootstrap 95\% CIs. (e) Both
    prior scores at one bias strength, in hops on the prior graph; bars are means, lines min--max
    over 10 seeds.}
  \label{fig:ablations}
\end{figure}

\subsection{Results}
\textbf{Coarse-graining costs almost nothing.} We first asked what the reduction costs. The
three reduced models from the first ablation study sit within half a point of one another, and the full panel is no better on
either metric ($17.1\%$/$14.8\%$; Figure~\ref{fig:ablations}a). The best fit of
the three is the 3-state control ($-0.4$ points against Co\textsuperscript{2}ODE, $p=0.004$),
which selects nothing and so recovers no context-specific structure. A ceiling in the data explains the flatness: a probe given the anchors,
context and strength predicts the target at $R^2=0.917$, and the other 27 markers add nothing
($0.913$). The
symbolic implementation, with a 1-dimensional latent instead of 4, compresses the 85-term
library to 3--4 terms per equation and sits $7$ points behind on the target
(Appendix~\ref{si:symbolic}).

\textbf{Prediction runs through the selected markers.} We next asked whether the model uses the
markers it selects. We gave a trained model another context's selected markers at inference, leaving its
context embedding and closure untouched. The median target error rises from $15.9\%$ to
$25.0\%$, and to $26.2\%$ with random markers (Figure~\ref{fig:ablations}b); the swap is worse
than the model's own markers in 36 of 38 contexts. The cost is the same in the closure-free
models, so the selection and not the latent carries the prediction. Selections also repeat across seeds: the Jaccard overlap of the three chosen markers is $0.56$, against $0.12$ for
mismatched contexts (permutation $p=2\times10^{-4}$; Figure~\ref{fig:ablations}c). For 19 of the 30 perturbations a single marker is kept in at least 8 of 10 seeds while rarely appearing elsewhere; no control has one.

\textbf{The closure buys robustness.} The closure encodes exactly the markers the coarse state excludes, so the model sees the whole panel either way and only the split between state and latent changes. An unusual split should then cost little. A seed that departs from the
most common selection predicts $4.1\%$ worse without the closure ($[1.0,7.4]$) and no worse with
it ($-2.3\%$, $[-6.9,2.7]$), at equal mean accuracy (Figure~\ref{fig:ablations}d). The closure
therefore makes error uniform across seeds rather than lower. The effect replicates in an
independent out-of-distribution test (Appendix~\ref{si:probes}).

\textbf{A prior rearranges the structure.} We last asked how tightly the data constrain the structure the model recovers, using each prior score as a probe.
Fit barely moves across the four bias settings: p-S6 rnMAE spans $16.1$--$16.5\%$ and the
three-marker average $14.1$--$14.5\%$. The structure moves instead. The node bias cuts the node
score from $1.06$ to $0.10$. It also moves the edge score, which it does not optimise, from
$2.85$ to $2.66$ ($p=0.002$ over 10 seeds), and both biases together reach $0.10$ and $2.63$
(Figure~\ref{fig:ablations}e). The edge bias alone moves neither score. A prior therefore
rearranges the recovered structure at no cost in fit.

\section{Discussion}

We introduce \textit{conditionally coarse-grained ODEs} (Co\textsuperscript{2}ODE) to decode rewiring in biochemical networks across different contexts. Our approach matches a 30-state model on the target, but so does a three-state control: on these data the reduction costs nothing and buys nothing. Its value is that the context-specific selection is explicit and reproducible across seeds where the full model leaves it implicit---and that the same machinery then shows how weakly these data constrain what it selects.
Indeed, adding a prior score to the loss moves the recovered structure onto known pathways at no cost
in accuracy. The fit is therefore near-indifferent between structures that disagree
on the prior graph, so agreement with known biology is not evidence that the structure is
right.

Deciding between the structures these data leave open needs interventions they do not contain.
We are designing a mass cytometry experiment after Lun \textit{et al.} (2019)~\cite{lun2019kinome},
adding an additional drug panel targeting phospho-proteins in the measured panel. Cutting an edge inside the
state breaks the correlations that make competing structures indistinguishable: the 3-state control
should then stop being sufficient on its own, and the markers this reduction discards should
start to matter.

\newpage

\begin{ack}
F.F. and T.P. were supported by the Francis Crick Institute, which receives its core funding
from Cancer Research UK (CC2242), the UK Medical Research Council (CC2242), and the
Wellcome Trust (CC2242). This work was funded by the European Union (ERC, DeepMechanism,
101163005). Views and opinions expressed are however those of the authors only and do not
necessarily reflect those of the European Union or the European Research Council Executive
Agency. F.F. reports consulting honoraria from Deep Origin, which had no influence on this study.
\end{ack}

\newpage
\appendix
\input{si_appendix.tex}

\end{document}

%% file: si_appendix.tex
\raggedbottom

\setlength{\floatsep}{20pt plus 4pt minus 2pt}
\setlength{\textfloatsep}{22pt plus 4pt minus 4pt}
\setlength{\intextsep}{20pt plus 4pt minus 2pt}

\makeatletter
\@ifpackageloaded{float}
  {\newenvironment{sitable}{\begin{table}[H]\centering}{\end{table}}}
  {\newenvironment{sitable}{\begin{table}[!ht]\centering}{\end{table}}}
\makeatother

\section{Data and preprocessing}
\label{si:data}

\paragraph{Split structure.} The split is random over trajectories, grouped by GFP bin and
seeded identically for every arm and every model seed. All 38 contexts therefore
appear in all three splits, and each trajectory appears in exactly one. Generalisation
reported here is across trajectories within a context; the architecture is conditioned on a
context embedding and does not transfer to a context it has never seen, so held-out-context
evaluation is not applicable.

\begin{sitable}
\small
\caption{Dataset and preprocessing. Unit of splitting is the trajectory.}
\begin{tabular}{@{}lp{10.6cm}@{}}
\toprule
Source        & Lun et al.\ (2019)~\cite{lun2019kinome}, single-cell mass cytometry \\
Preprocessing & GP-smoothed trajectories, GFP-binned, with a derived strength column \\
Panel         & 30 phospho-markers \\
Conditions    & 40 total; 38 retained: 30 single-gene overexpressions and 8 controls \\
Dropped       & 2 contexts whose time grids do not match the rest (DUSP4, MAP4K2) \\
Stimulus      & EGF, upstream of the network \\
Time          & snapshots over 1\,h post-stimulation \\
Smoothing     & Gaussian-process fit per context~\cite{rasmussen2006gaussian} \\
Target        & p-S6 \\
Anchors       & p-MEK1-2, p-AKT (fixed across contexts, always retained) \\
Strength $s$  & GFP-derived induction proxy, scaled to $[0,1]$, predicted rather than observed \\
Split         & 80/10/10 by trajectory: 2919 / 365 / 365 \\
Split seed    & fixed at 42 across every arm and every model seed \\
\bottomrule
\end{tabular}

\end{sitable}

\section{Architecture and hyperparameters}
\label{si:hparams}\label{sec:appendix}

Base preset, with the
per-arm overrides in Table~\ref{tab:si-arms}.

\begin{sitable}
\small
\caption{Architecture and optimisation, identical across arms except where
Table~\ref{tab:si-arms} states otherwise. HSIC is the Hilbert--Schmidt independence criterion~\cite{gretton2005measuring}.}
\begin{tabular}{@{}llp{6.2cm}@{}}
\toprule
Component & Setting & Value \\
\midrule
Selector & conditioning dim & 16 \\
         & hidden & 128 \\
         & number of conditions & 38 \\
         & relaxation temperature & annealed $2.5\rightarrow0.6$, rate $0.95$, from epoch 10 \\
         & Gumbel sampling & off (straight-through top-$k$) \\
         & explicit slots & $6 = 3$ fixed (target $+$ 2 anchors) $+\ k{=}3$ learned \\
\addlinespace
Conditional VAE & encoder hidden & $(128, 64)$, dropout $0.15$ \\
         & decoder hidden & $(64, 64)$, dropout $0.15$ \\
         & $z_{\perp}$ (adversarial) & 4 \\
         & $z_{\parallel}$ (strength-supervised) & 3 \\
         & gradient-reversal scale & 3 \\
\addlinespace
Dynamics & type & MLP, hidden 192 \\
         & predictor hidden & 192 \\
         & context embedding dim & 32 \\
         & solver & RK4, fixed step \\
         & substeps per observed interval & 2 \\
\addlinespace
Optimiser & Adam, learning rate & $1.622\times10^{-3}$ \\
          & weight decay & $2.542\times10^{-7}$ \\
          & gradient clip & $0.5$ \\
          & batch size & 128 \\
          & lr schedule & 10 warmup epochs, min scale $0.5$ \\
          & epochs / early stop & 400 / 50 \\
\addlinespace
Loss weights & $\lambda_{\mathrm{rec}}$, $\lambda_{s}$ & $1.0$, $1.0$ \\
             & $\lambda_{\mathrm{adv}}$ & $0.1018$ (trajectory $0.2607$), warmup 10 epochs from $0.2$ \\
             & $\beta_{\mathrm{KL}}$ & $0.02$ \\
             & $\lambda_{\mathrm{sp}}$ & swept; see Table~\ref{tab:si-sweep} \\
             & latent invariance, correlation / HSIC & $2.23{\times}10^{-3}$ / $4.59{\times}10^{-3}$ \\
             & latent trajectory, corr / HSIC & $4.14{\times}10^{-2}$ / $8.28{\times}10^{-2}$ \\
             & latent L2 / velocity & $4.64{\times}10^{-5}$ / $3.42{\times}10^{-5}$ \\
             & trajectory latent supervision & $0.5$ \\
             & fit schedule & mix over 60 epochs, $\lambda\ 0.5\rightarrow0.2$ \\
\bottomrule
\end{tabular}

\end{sitable}

\begin{sitable}
\small
\caption{The four arms of the state-dimension ablation.}
\label{tab:si-arms}
\begin{tabular}{@{}lccp{7.2cm}@{}}
\toprule
Arm & State dim & Latent in ODE & Composition of the state \\
\midrule
3-state & 3 & no & target and 2 anchors only; no marker selection and no latent closure \\
\addlinespace
6-state & 6 & no & target, anchors, and the 3 selector-chosen markers; no latent closure \\
\addlinespace
Co\textsuperscript{2}ODE & 10 & yes, $\dim 4$ & the 6 markers above plus the 4-dimensional $z_{\perp}$ evolved inside the ODE \\
\addlinespace
Full panel & 30 & --- & all 30 markers as ODE states, context-conditioned, no selection or closure \\
\bottomrule
\end{tabular}

\end{sitable}

\section{Metric}
\label{si:metric}

Range-normalised MAE (rnMAE) is computed per trajectory and per marker:
\begin{equation}
  \mathrm{rnMAE} \;=\; 100 \times
  \frac{\frac{1}{T}\sum_{t}\lvert \hat y_t - y_t\rvert}
       {\max\!\big(\max_t y_t - \min_t y_t,\ \delta\big)},
  \qquad \delta = 0.5 .
\end{equation}
The floor $\delta$ prevents blow-up on degenerate flat trajectories. The floor does engage. The median test-trajectory range for p-S6 is $0.79$, and $27\%$ of test
trajectories fall below $\delta$. On that subset the denominator is the floor, not the
trajectory's own range, which compresses the differences between all models. Trajectories are integrated from $t=0$; no
teacher forcing is used at evaluation. Cells are aggregated by averaging per-trajectory rnMAE,
never by pooling absolute errors and dividing by a pooled range. ``mean3'' is the unweighted
mean of rnMAE over $\{$p-S6, p-MEK1-2, p-AKT$\}$.

\paragraph{Aggregation.} Figure~\ref{fig:ablations}a,b
box the 38 per-context values, each a median over model seeds. Numbers quoted in
Section~\ref{sec:experiments} use the same reduction: rnMAE is first reduced per context as
the median over model seeds, and the reported figure is the median over the 38 contexts.
The two statistics will not agree to the decimal; both are given so the figure and the text can
be read against each other.

\paragraph{Trivial baselines.} rnMAE is relative, so its scale needs anchoring. These rows give what it
returns for predictors that model no dynamics at all, on the same test split, target and
aggregation as every other number in the paper.

\begin{sitable}
\small
\caption{What rnMAE gives for predictors that learn nothing, on p-S6. ``Best flat line'' is the
constant minimising absolute error for each trajectory, chosen in hindsight, so it is a lower
bound on any constant prediction and not available to a causal predictor. The learnt arms are
included for scale.}
\label{tab:si-baselines}
\begin{tabular}{@{}lr@{}}
\toprule
Predictor & rnMAE p-S6 (\%) \\
\midrule
Global mean trajectory & 86.0 \\
Linear extrapolation from $t_0,t_1$ & 37.1 \\
Hold the initial value & 36.8 \\
Per-context mean trajectory & 36.7 \\
Best flat line per trajectory, fitted in hindsight & 28.4 \\
\midrule
Symbolic variant, 3--4 terms per equation (Appendix~\ref{si:symbolic}) & 23.4 \\
Co\textsuperscript{2}ODE & 16.0 \\
\bottomrule
\end{tabular}
\end{sitable}

The per-context mean trajectory is the context lookup table: it is given the context identity
and returns that context's average behaviour, with no dependence on the initial state. It
reaches $36.7\%$. Co\textsuperscript{2}ODE reaches $16.0\%$, so the reduced model is not
reproducing a per-context average, and the flatness of the state ladder in
Figure~\ref{fig:ablations}a is not the flatness of a metric that cannot separate anything.

\section{Training, sweeps, and model selection}
\label{si:selection}

\paragraph{Selection metric.} Early stopping and best-checkpoint selection use a
\emph{reporters-only} validation metric, computed on the target and anchors alone rather than
uniformly over all state slots. With a uniform-over-slots metric the reporter share
of the selection signal shrinks as the state grows (100\% at 3 states, ${\sim}50\%$ at 6,
${\sim}10\%$ at 30), so larger arms have their checkpoint chosen on markers the paper never
forecasts. Every number reported here uses reporters-only selection. For the 3-state arm the two
metrics coincide, since all three of its slots are reporters, so that arm is an unchanged
control.

\paragraph{Sweep.} Both nuisance hyperparameters are swept identically for every arm and
selected per arm on validation loss; test is reported once.

\begin{sitable}
\small
\caption{Sweep grid, applied identically to the three reduced arms ($3\times3\times10 = 90$ runs each); the full panel had its own width and learning-rate sweep.}
\label{tab:si-sweep}
\begin{tabular}{@{}ll@{}}
\toprule
Swept quantity & Grid \\
\midrule
Reporter weighting of the trajectory loss & $\{0.5,\ 0.7,\ 0.9\}$ \\
Jacobian group-sparsity strength $\lambda_{\mathrm{sp}}$ & $\{0,\ 1{\times}10^{-2},\ 3{\times}10^{-2}\}$ \\
Model seeds & $0 \ldots 9$ \\
\bottomrule
\end{tabular}

\end{sitable}

\paragraph{Full-panel baseline tuning.} The full panel received the same
reporter-weighted objective and reporters-only selection, the same training budget
(400 epochs / early stop 50), a width sweep (hidden $\in\{192,384,768\}$) and a
learning-rate sweep ($\{8{\times}10^{-4},\ 1.6{\times}10^{-3}\}$). Hidden 192 with lr
$8{\times}10^{-4}$ won; \emph{more} capacity made it worse (hidden 768 reached a validation
training-objective loss of 0.222 against 0.161 at 192; this is the training loss, not rnMAE), so the baseline is not capacity-limited.

\section{Prior graph and the two scores}
\label{si:prior}

\begin{sitable}
\small
\caption{Prior interaction network.}
\begin{tabular}{@{}lp{10.6cm}@{}}
\toprule
Resource           & OmniPath~\cite{turei2016omnipath}, directed literature-curated interactions \\
Access             & OmniPath \texttt{enzsub} endpoint via the \texttt{omnipath} Python client, downloaded 28 April 2026 \\
Layer and filters  & the enzyme--substrate (\texttt{enzsub}) layer, human, restricted to
                     modifications containing ``phosphorylation'', taken directed; unsigned
                     (activation and inhibition are not distinguished); self-edges dropped.
                     Retained edges carry curation effort 3--44 (median 4) \\
Graph size         & 70 directed edges over 28 of the 30 panel markers; the remaining 2 have no
                     retained edge and are unreachable by construction \\
Marker mapping     & curated per-marker gene-symbol sets, so a phospho-marker maps to every
                     gene of its family (p-RAF $\rightarrow$ ARAF, BRAF, RAF1; p-AKT $\rightarrow$
                     AKT1--3; p-ERK1-2 $\rightarrow$ MAPK1, MAPK3), and an edge is kept only when
                     both endpoints land on panel markers \\
Cost basis         & pairwise shortest-path distance on the directed graph \\
Latent coordinates & treated as \texttt{unreachable\_cost} (no graph identity) \\
Self loops         & \texttt{self\_loop\_cost}, set separately from the distance lookup \\
\bottomrule
\end{tabular}

\end{sitable}

\paragraph{Node score.} Mean prior cost of the three markers the selector chooses, anchors and
target excluded, scored by position
on the shortest path from the intervened gene to the target: on a short route scores low
(good), unreachable scores at the ceiling. The 8 control contexts have no intervened gene,
so the node score is defined only for the 30 overexpressions and the prior-score results are
over those.

\paragraph{Edge score.} Mean prior cost of the dependencies the dynamics express --- a nonzero
Jacobian entry $\partial f_{c,i}/\partial u_j$ or an active symbolic term --- scored by the
shortest-path distance from $j$ to $i$. A direct edge costs 0; each additional hop beyond
\texttt{free\_hops} costs more; an unconnectable pair costs \texttt{unreachable\_cost}
(shortest-path costs precomputed once per context).

\paragraph{Prior scores by bias setting.} Node score, then edge score, means over 10 seeds,
in hops on the prior graph: none $1.06$ / $2.85$; node bias $0.10$ / $2.66$; edge bias $1.09$ /
$2.87$; both $0.10$ / $2.63$. The edge bias moves the edge score only at a thirty-fold stronger
setting ($2.70$, $p=0.004$). That setting is a separate sweep at 4 substeps rather than 2, so it
is compared only against its own unbiased control, against which it costs $+0.8\%$ of validation
loss over 10 seeds ($p=0.52$); it is not comparable with the four matched-strength settings above. Fit across the same four settings spans $16.1$--$16.5\%$ on p-S6
and $14.1$--$14.5\%$ on mean3.

\paragraph{Where the scores are evaluated.}
When a bias is added to the loss it uses the same cost matrix, so reporting the biased model's
own penalty would be circular. Figure~\ref{fig:ablations}e therefore scores every arm,
biased and unbiased, with the identical unbiased graph cost.

\section{Probe and ablation protocols}
\label{si:probes}

Table~\ref{tab:si-protocols} gives, for each claim in Section~\ref{sec:experiments} that does not
come from the fit table, what was measured and the value obtained. All probes are fitted on the
training split and evaluated on the held-out test split.

\begin{sitable}
\small
\caption{Protocols behind each structural claim.}
\label{tab:si-protocols}
\begin{tabular}{@{}p{3.1cm}p{4.7cm}p{5.1cm}@{}}
\toprule
Protocol & What is done & Reported \\
\midrule
Conditioned completeness & Predict p-S6 at $t{=}1\ldots5$ from $t{=}0$. The probe is a random-forest regressor (300 trees, minimum leaf size 2) predicting all five future timepoints jointly, scored by 5-fold cross-validated $R^2$ averaged over outputs; context enters one-hot and strength as a scalar. A linear probe understates the anchors and so makes the panel look informative; the result below requires a nonlinear probe & $R^2 = 0.796$ anchors only; $0.917$ adding context and strength; $0.913$ full panel with conditioning \\
\addlinespace
Random-triple control & Selector's markers against random marker triples, under the completeness probe & $0.883$ against $0.889$: the selected markers are no more informative than random ones. The marker-swap test asks the different question of whether the trained model uses them \\
\addlinespace
Marker swap & At inference, give a trained model another context's markers, leaving its context embedding and latent closure untouched & p-S6 rnMAE rises from $15.9$ to $25.0$ (median over contexts), and the swap is worse than the model's own markers in 36 of 38 contexts; random markers are worse in 38 of 38 \\
\addlinespace
Off-consensus cost & Split runs by whether their selection matches the most common one for that context, and compare error within each group & $+4.1\%$ $[1.0, 7.4]$ closure-free against $-2.3\%$ $[-6.9, 2.7]$ with the closure; mean error $14.39$ against $14.44$ \\
\addlinespace
Selector-variability cost & Rank-correlate a run's selection agreement with the most common selection against its error, per (seed, context); 300 points per arm, bootstrap CIs clustered over seeds; on the p-S6/MEK/AKT average. Percentages are relative error changes; rnMAE differences elsewhere are in points & closure-free $\rho = -0.134$ $[-0.186, -0.084]$; closure $-0.000$ $[-0.131, +0.105]$. Replicated in an independent strength-OOD grid (closure-free $+11.8\%$ $[+5.1, +18.3]$; closure $-3.8\%$ $[-10.7, +3.2]$). Pinning the closure arm's selections into the closure-free model costs $+0.51$ points $[-0.54, +1.54]$ against a peer seed's, so the picks are different, not worse \\
Strength and context probe & Predict $s$ and context identity from $z_{\perp}$, against the same from the selected markers. Continuous targets: standardised ridge regression, penalty chosen by generalised cross-validation over 13 log-spaced values, scored by 5-fold $R^2$. Categorical targets: standardised multinomial logistic regression, scored by 5-fold cross-validated accuracy & $R^2 = 0.71$ for strength; $0.20$ against $0.83$ for context identity on the encoder's sample, $0.80$--$0.98$ once evolved in the ODE (chance $0.03$) \\
\addlinespace
Seed reproducibility & Overlap of selected free-marker sets for the same context across seeds, against a mismatched-context null & $0.56$ against $0.12$ ($4.7\times$); permutation $p = 2\times10^{-4}$; 36--37 distinct supports of 38 \\
\addlinespace
Random-support retrain & Retrain with a random support of equal size & $p = 0.026$ \\
\addlinespace
Full-panel context specificity & Cross-context cosine of non-anchor dependency patterns & $0.997$ over 5 seeds: one pattern for all 38 contexts \\
\bottomrule
\end{tabular}

\end{sitable}

\section{Statistical treatment}
\label{si:stats}

\paragraph{Unit of analysis.} Each arm is trained with 10 seeds (full-panel control: 5). rnMAE
is first reduced per context, as the mean over the test trajectories of that context, giving 38
values per seed. Comparisons between arms are therefore paired: the same 38 contexts and the
same test split are seen by every arm.

\paragraph{Test.} Arms are compared by a two-sided Wilcoxon signed-rank test over the 38
paired per-context values. We report the mean paired difference, the $p$-value, and the number of contexts favouring each
arm, since with 38 paired samples and a consistent sign a difference of a few tenths of a point
is detectable while remaining small: significance and magnitude are reported separately and
should be read separately.

\begin{sitable}
\small
\caption{The state ladder, medians over the 38 contexts. These are the levels behind the paired
differences in the next table, and behind ``within half a point'' in
Section~\ref{sec:experiments}.}
\label{tab:si-ladder}
\begin{tabular}{@{}lrr@{}}
\toprule
Arm & p-S6 (\%) & mean3 (\%) \\
\midrule
3-state control & 15.6 & 13.8 \\
6-state (closure-free) & 16.1 & 14.0 \\
Co\textsuperscript{2}ODE (10-state) & 16.0 & 14.3 \\
Full panel (30-state) & 17.1 & 14.8 \\
\bottomrule
\end{tabular}
\end{sitable}

\begin{sitable}
\small
\caption{Paired comparisons of p-S6 rnMAE over the 38 contexts, all arms at the matched
protocol. On the three-marker average the same comparisons give $-0.23$ ($p = 0.005$),
$-0.11$ ($p = 0.194$), $-0.61$ ($p = 0.046$) and $-0.85$ ($p = 0.002$): Co\textsuperscript{2}ODE
separates from the full panel on mean3 but not on the target alone.}
\begin{tabular}{@{}lrrp{3.4cm}@{}}
\toprule
Comparison, p-S6 & $\Delta$ mean & Wilcoxon $p$ & Contexts favouring smaller \\
\midrule
3-state against Co\textsuperscript{2}ODE  & $-0.44$ & $0.004$ & 29/38 \\
6-state against Co\textsuperscript{2}ODE  & $-0.23$ & $0.243$ & 22/38 \\
Co\textsuperscript{2}ODE against full panel & $-0.33$ & $0.243$ & 22/38 \\
3-state against full panel & $-0.77$ & $0.005$ & 29/38 \\
\bottomrule
\end{tabular}

\end{sitable}

\paragraph{Inductive-bias arms.} The four bias settings are compared to the unbiased arm by the
same paired test over the 38 contexts, at the matched protocol and with 10 seeds each. On p-S6
the node bias gives $-0.32$ ($p = 0.005$), the edge bias $-0.19$ ($p = 0.040$) and both together
$-0.22$ ($p = 0.118$); on the three-marker average none of the three separates
($p = 0.171$, $0.966$, $0.099$). The biases therefore do not cost fit --- if anything they
improve it slightly --- but the effect is ${\sim}2\%$ of the error, against a $90\%$ reduction in
the node score, so the objective is close to indifferent between structures that differ
wholesale on the prior graph.

\paragraph{Whole-state loss.} The masked trajectory loss
differs by ${\sim}36\%$ between the unbiased and node-biased arms (median $0.136$ against
$0.087$), but this is not a like-for-like comparison: the bias changes which markers occupy the
state, so the two arms are scored on different slots. Only the reporter metrics, which are fixed
by construction, are comparable across arms, and those are the ones reported.

\paragraph{Structural claims.} The seed-reproducibility result uses a permutation test against
a mismatched-context null ($p = 2\times10^{-4}$, $10^{4}$ permutations); the random-support
retrain comparison uses a paired test over contexts ($p = 0.026$); the out-of-distribution
comparison reports per-protocol paired tests ($p = 0.093$ on p-S6, $0.013$ on mean3 for the
hardest protocol). Probe $R^2$ and accuracy values are single point estimates on the held-out
split and are reported without intervals.

\paragraph{Multiplicity.} All tests are two-sided and treat the context as the unit of
analysis. Reported $p$-values are unadjusted: the comparisons are pre-specified but
exploratory. Several claims we report sit between $0.01$ and $0.05$ and should be read as
such.

\section{The symbolic variant}
\label{si:symbolic}

The dynamics module has two interchangeable implementations (Section~\ref{sec:dynamics}). Every
result in the main text uses the sparse neural one. This section reports the conditional SINDy
variant, which returns the recovered structure as equations rather than as Jacobian entries.

The symbolic runs use a different Co\textsuperscript{2}ODE configuration. The 6-marker state is
the same: the target, the two anchors, and the 3 selected markers. The latent is 1-dimensional
rather than the 4 of Table~\ref{tab:si-arms}, so the ODE has 7 states rather than 10. Each
latent state adds an equation and its couplings to the library, which makes the coefficients
harder to read. The gap to the neural numbers is therefore partly configuration and not only
the dynamics module. The latent enters the target's equation in 0--6 of the 38 contexts.

All 260 symbolic runs use the same data, split and metric as the main text. The table below
rests on the 205 that are compatible with the current implementation and do not diverge,
spanning 72 configurations at 1 to 4 seeds each. Fifteen runs diverge and are excluded; a further 40 belong
to an earlier version of the dynamics module and are not comparable with it. That second
exclusion is not random: those runs are the only ones trained without a learned selector, so
the symbolic arm has no selector ablation, and its results should be read as conditional on the
selector rather than as a test of it.

\begin{sitable}
\small
\caption{The symbolic variant, 205 non-diverged runs grouped by active terms per equation,
on the same metric and split as the neural arms at the foot of the table; Table~\ref{tab:si-baselines} gives the trivial baselines for the same metric. Both error columns are
medians over the 38 contexts and then over runs, with the range across runs in brackets; mean3
averages the target and the two anchors. Compressing an 85-term library to $3$--$4$ active terms
costs $2.2$ points of target error against the densest runs and $1.3$ points of mean3 against the
best band, so the reduction is inexpensive but not free. The neural rows are the state-ladder sweep of
Figure~\ref{fig:ablations}a.}
\label{tab:si-symbolic}
\begin{tabular}{@{}lrrrrr@{}}
\toprule
Model & Runs & Terms/eq. & p-S6 (\%) & mean3 (\%) & Target eq.\ non-empty \\
\midrule
Symbolic, 3--4 terms & 62 & 3.5 & 23.4 [21.9, 27.4] & 23.5 [21.7, 26.4] & 38/38 \\
Symbolic, 4--8 terms & 45 & 4.5 & 23.2 [20.8, 30.5] & 23.3 [20.9, 28.3] & 37/38 \\
Symbolic, 8--20 terms & 36 & 11.8 & 25.5 [21.3, 32.4] & 23.9 [21.3, 30.0] & 36/38 \\
Symbolic, 20--60 terms & 45 & 23.4 & 23.3 [19.9, 25.9] & 22.2 [18.1, 24.6] & 38/38 \\
Symbolic, 60--85 terms & 17 & 85.0 & 21.2 [19.7, 21.5] & 23.4 [19.9, 23.6] & 38/38 \\
\midrule
Neural, Co\textsuperscript{2}ODE & & & 16.0 & 14.3 & \\
Neural, full panel & & & 17.1 & 14.8 & \\
\bottomrule
\end{tabular}
\end{sitable}

Compression is cheap here too, though its cost depends on the metric. On the
target, going from $20$--$60$ active terms to
$3$--$4$ costs $0.1$ points (Mann--Whitney $p=0.03$ on a $0.1$-point difference) and only runs
keeping essentially the whole library are meaningfully better, by $2.2$; the intermediate
bands differ by $0.1$--$0.2$. On the three-marker average the
ordering differs: the $20$--$60$ band is best at $22.2\%$, compressing to $3$--$4$ terms costs
$1.3$ points ($p<0.001$), and the densest runs are no better than the sparsest. Neither metric shows error
growing steadily with sparsity: among models under 20 terms, the correlation between terms and
target error is $+0.24$ ($n=143$, $p=0.005$). The symbolic variant is behind the
neural one throughout, $23.4\%$ against $16.0\%$ on the target, which is why the main results use
the neural implementation.

The configuration we would report is the cap-scheduled, graph-biased one. Three of its four
seeds reach $5.6$--$5.8$ terms per equation at $22.6$--$23.1\%$ rnMAE, with the target's own
equation non-empty in all 38 contexts and carrying a numerically non-negligible coupling in 37
of them. For the ALPK2 context, seed 0 returns
\begin{align*}
\dot{u}_{\mathrm{S6}} =\ & (0.040 - 0.009 s) - (0.017 + 0.012 s)\,u_{\mathrm{S6}}
  + (-0.008 + 0.012 s)\,u_{\mathrm{AKT}} \\
  & - (0.010 + 0.024 s)\,u_{\mathrm{S6}}^{2}
  + (0.005 - 0.006 s)\,u_{\mathrm{AKT}} u_{\mathrm{JNK}},
\end{align*}
with the induction strength $s$ modulating every coefficient. Across the 38 contexts the
target's equation couples most often to the two anchors, p-MEK1-2 in 15 contexts and p-AKT in
11. Both anchors are in the state by construction in every context, so this is a statement
about which couplings the sparsifier keeps, not about which markers were available.

Sparsity comes from the prior graph and the cap schedule, not from the sparsity penalty: the
unstructured proximal sparsifiers (L1, elastic net, group L1) sit at 62--85 terms per
equation whatever their penalty strength, which is no reduction against the library.

Three failure modes qualify this. Fifteen of the 220 evaluated runs diverge outright
(test trajectory MSE above 10, against roughly $0.12$ for healthy runs) and are excluded above;
in the reported configuration that is one seed in four. Sparsity can also come from emptying the target's equation
rather than compressing it: the most aggressive second-pass sparsifier reaches $3.5$ terms per
equation but reduces the target to a strength-modulated constant in 11 of 38 contexts. And the
accuracy-capped configurations delete the target's equation outright in 4 of 38 contexts. A term
count alone therefore cannot distinguish compression from deletion, and we report the count of
contexts with a surviving, non-negligible target equation alongside it.

\section{Compute}
\label{si:compute}

\begin{sitable}
\small
\caption{Compute. The dominant cost is the identical hyperparameter sweep applied to every arm
rather than any single model: one Co\textsuperscript{2}ODE model trains in under an hour on 8
cores.}
\begin{tabular}{@{}p{4.6cm}p{8.8cm}@{}}
\toprule
Hardware & CPU only; no GPU used at any stage \\
Per run  & 8 CPU cores, 24\,GB RAM \\
Wall-clock, Co\textsuperscript{2}ODE and reduced controls & ${\approx}50$\,min per run, ${\approx}6.5$ CPU-hours \\
Wall-clock, full-panel control & ${\approx}30$\,min per run, ${\approx}4$ CPU-hours \\
Reported state-dimension experiment & 270 sweep runs $+$ 29 of 30 full-panel runs (one failed to finish), ${\approx}1900$ CPU-hours \\
All experiments including exploratory & ${\approx}2700$ CPU-hours \\
Scheduler & Slurm, shared institutional CPU partition \\
\bottomrule
\end{tabular}

\end{sitable}

\section{Code, data, and licences}
\label{si:assets}

\paragraph{Code.} Code is not released for review, and will be released on publication.
Section~\ref{si:repro} gives the exact commands and Section~\ref{si:hparams} the full
hyperparameter set, so the results are reproducible from the description alone.

\paragraph{Data.} Single-cell mass-cytometry measurements from Lun \textit{et al.}
(2019)~\cite{lun2019kinome}, obtained from the resource published with that paper
(Mendeley Data, \texttt{10.17632/3kh7ypz232.1}). We use it unmodified apart from the preprocessing in
Section~\ref{si:data}. The dataset contains no human-subject or personally identifying
information: it comprises measurements on a cultured cell line.

\paragraph{Prior network.} OmniPath~\cite{turei2016omnipath}, \texttt{enzsub} endpoint, accessed 28 April 2026. OmniPath
aggregates literature-curated resources that retain their original licences; we use it for
scoring only and redistribute no part of it.

\paragraph{Software.} PyTorch 2.8.0, NumPy 1.26.4, SciPy 1.16.1, pandas 2.3.1,
scikit-learn 1.7.1, matplotlib 3.10.5, NetworkX 3.5 and the \texttt{omnipath} client --- all under BSD-, MIT- or
Apache-family licences and used as published. The conditional SINDy module is our own
implementation of the method of Brunton et al.~\cite{brunton2016discovering}; no
symbolic-regression package is used in any reported run.

\section{Exact run configuration}
\label{si:repro}

Sections~\ref{si:hparams}--\ref{si:selection} state the configuration in full; the invocation
below is included only so that it is unambiguous which settings were non-default. One
Co\textsuperscript{2}ODE seed:
{\footnotesize
\begin{verbatim}
python -m train --model_name PRESET \
  --data_path data/experimental/lun_2019_100gfp_with_gfpbin_strength_gpfit.csv \
  --perturbation_col scenario --strength_col strength \
  --val_size 0.1 --test_size 0.1 --split_random_state 42 --seed 0 \
  --drop_mismatched_scenarios --no-explicit_inputs_as_exogenous --conditional_vae \
  --explicit_input_features "p-AKT,p-MEK1-2" --explicit_target_features "p-S6" \
  --strength_source pred --max_explicit 6 --evolve_adv_in_ode --latent_adv 4 \
  --lambda_neural_jacobian_l21 3e-2 --traj_reporter_frac 0.9 --select_on_reporters \
  --beta_kl 0.02 --epochs 400 --early_stop 50 --device cpu --output_dir OUT
\end{verbatim}}
The reduced controls differ only in the state composition given in
Table~\ref{tab:si-arms}, with the latent closure disabled. The full-panel control is a
separate 30-state model, context-conditioned, at hidden width 192 and learning rate
$8{\times}10^{-4}$ (the winner of its own width and learning-rate sweep), with the same
group-sparsity penalty and the same reporters-only selection.
\section{Limitations}
\label{si:limits}

\paragraph{Out-of-distribution behaviour.} Across four strength-extrapolation protocols
Co\textsuperscript{2}ODE is worse than the 3-state control on mean3 in all four; p-S6 is mixed
and never significant. On the hardest protocol, with dose bins 85--99 held out and validation
on 70--84: 3-state $30.64$ / $34.91$ against Co\textsuperscript{2}ODE $32.17$ / $36.97$
($p = 0.093$ / $0.013$).

\paragraph{Coupling-level reproducibility.} The overlap of $0.56$ in Section~\ref{sec:experiments} is over selected marker sets. The recovered couplings themselves do not reproduce across seeds: pairwise agreement of the dependency support is $0.032$ closure-free and $0.036$ with the closure, both at chance. Both arms behave the same way, so this is a property of the Jacobian readout. Which markers a context uses repeats. Which couplings the model writes down among them does not.

\paragraph{Latent leakage.} The adversarial head acts on the encoder's sample; no penalty in the reported
runs acts on the evolved latent. A sweep that does, $\lambda_{\mathrm{inv,ode}}\in
\{0.03,0.1,0.3,1.0\}$ at 10 seeds each, reduces the strength correlation of the evolved
latent from $0.61$ to $0.47$ at $\lambda=0.3$, with validation fit slightly better
($0.0291$ against $0.0311$). The leak is reduced, not removed.

\paragraph{Sparsity.} No sparsity advantage is claimed. We measure concentration of the
target's Jacobian row by $k_{90}$, the number of entries, ranked by absolute value, needed to
reach $90\%$ of the row's absolute mass. Within an arm the L21 penalty changes accuracy
($\Delta$rnMAE at $\lambda = 3{\times}10^{-2}$: 3-state $+0.72$, 6-state $+0.35$,
Co\textsuperscript{2}ODE $-0.51$) but does not reduce $k_{90}$, so it regularises rather than
prunes. We make no cross-arm comparison, since $k_{90}$ is a raw count and a wider state can
spread the same mass over more entries. The full-panel control carries the same penalty, so a
sparsity-penalised 30-state model of the kind proposed in~\cite{aliee2022sparsity} is already in
the comparison and fits no better (Table~\ref{tab:si-ladder}); we do not report its $k_{90}$, so
whether that penalty leaves a 30-state structure readable is open.

\paragraph{Discretisation.} Results are reported at the training discretisation (RK4, 2
substeps per observed interval), and every arm was trained at that setting with no teacher
forcing. Evaluating a trained model under a finer integrator degrades fit and shifts the
recovered dependencies, but the integrator is part of the fitted model, so this measures
sensitivity to swapping it and not convergence of a vector field. We do not run the convergence
test, which would retrain across substep counts and compare the recovered structure.
Comparisons here are therefore valid between runs at matched substeps. Runs at 4 substeps exist
for the reduced arms and are excluded, since no full-panel counterpart was trained that way.

\paragraph{Seed counts.} Co\textsuperscript{2}ODE and the reduced controls use 10
seeds each; the full-panel baseline uses 5. The baseline's spread is small relative to the
gap: test-set $R^2$ ranges over $0.936$--$0.941$ across its seeds.